\documentclass[12pt]{iopart}

\usepackage{mathrsfs}

\usepackage{graphicx}
\usepackage{psfrag}
\usepackage{stackengine}
\usepackage{braket}

\usepackage{iopams}
\usepackage{slashed}

\expandafter\let\csname equation*\endcsname\relax

\expandafter\let\csname endequation*\endcsname\relax

\usepackage{amsmath}

\usepackage{cite}

\newcommand{\To}{\rightarrow}

\newcommand{\be}{\begin{equation}}
\newcommand{\ee}{\end{equation}}
\newcommand{\bea}{\begin{eqnarray}}
\newcommand{\eea}{\end{eqnarray}}
\newcommand{\qqq}{\end{document}}
\newcommand{\bspl}{\begin{split}}
\newcommand{\espl}{\end{split}}

\newcommand{\bV}{\begin{pmatrix}}
\newcommand{\eV}{\end{pmatrix}}
\newcommand{\de}{\partial}

\newcommand{\DE}{\overset{\centerdot}}
 \newcommand{\DEE}{\overset{\centerdot\centerdot}}

\newcommand{\s}{\mathscr{S}}

\begin{document}
\title{One-dimensional Bose gas driven by a slow time-dependent harmonic trap}
\author{Stefano Scopa and Dragi Karevski}
\address{Institut Jean Lamour, dpt. P2M, Groupe de Physique Statistique, Universit\'e de Lorraine, CNRS UMR 7198, B.P. 70239, F-54506 Vandoeuvre les Nancy Cedex, France}
\ead{\mailto{dragi.karevski@univ-lorraine.fr}}

\begin{abstract}
We consider  the unitary time  evolution of a one-dimensional cloud of hard-core bosons loaded on a harmonic trap potential which is slowly released in time with a general ramp $g(t)$. After the identification of a typical length scale $\ell(t)$,  related to the time ramp, we focus our attention on the dynamics of the density profile within a first order time-dependent perturbation scheme. In the special case of a linear ramp, we compare the first order predictions to the exact solution obtained through Ermakov-Lewis dynamical invariants. 
We also obtain an exact analytical solution for a cloud released from a harmonic trap with an amplitude that varies as the inverse of time.  In such situation, the typical size of the cloud grows with a power law governed by an exponent that depends continuously on the initial trap frequency. At high enough initial trap amplitude, the exponent acquires an imaginary part that leads to the emergence of a 
log-periodic modulation of the cloud expansion. 
\end{abstract}

\section{Introduction}
Advances in ultracold atomic gases have led to the possibility of realizing low-dimensional model systems, e.g.\cite{13,14,15,16,17,18,19}. This has opened the road to probe experimentally theoretical predictions on the dynamical aspects related to non-equilibrium effects, see \cite{QKZ,polkovnikov} for recent reviews. 
 In particular, quantum gases parametrically driven through a quantum phase transition have played a central role \cite{QKZ} for testing new ideas related to the thermalization (or the absence of it) of such gases. A paradigmatic model in this context is the Bose-Hubbard (BH) model \cite{BH1,BH2}, describing interacting bosons on a lattice, and exhibiting  Superfluid-to-Mott Insulating  (SF-MI) quantum phase transitions  \cite{BH2,Sachdev}.
To mention just a few examples of the wide literature dedicated to such so called quantum quenches, we may quote J. M. Zhang et al. \cite{U1,U2} that have studied the quantum quench driven by the on-site interaction or J. Dziarmaga et al. \cite{JDZ1,JDZ2,J3}  that have investigated the case of a time-dependent hopping magnitude, focusing on the loss of adiabaticity and on Kibble-Zurek scaling regims \cite{KZ1,KZ2}. 

In this context, our aim here is to investigate close to the SF-MI transition the dynamics of a one dimensional cloud of bosons driven by the slow release of an inhomogeneous confining potential. We focus our attention on the case of a harmonic potential $V(x)\sim g(t) x^2$ for a given time ramp $g(t)$.  We consider in particular the low-density regime of the Bose gas in the limit of high repulsive interactions for which the Tonks-Girardeau model \cite{LL,TG1,TG2,TG3} is a good effective description.
On one hand this problem can be handled within the framework of quantum quenches ideas where off-equilibrium behaviors are related to the breakdown of adiabaticity near the phase transition. 
The dynamical behavior after a sudden release of the trap has been studied so far quite extensively, see for example \cite{CARK,CSC1,CSC2,MAZ}.
In the case of a slow driving, which is the situation considered in this work, the system can be investigated perturbatively around the adiabatic evolution \cite{polkovnikov}. The departure from the equilibrium can be quantified studying the first off-equilibrium corrections of the density profile or counting, {\it \`a la} Kibble-Zurek, the number of excitations generated during the quench \cite{nonlin}.
On the other hand, another way to approach the problem is to find dynamical invariants associated to the time-dependent system, which basically after diagonalization reduces to a set of time-dependent harmonic oscillators, see e.g. \cite{EL1,EL2}. For the special case of a linear ramp $g(t)$ one is then able to solve explicitly the associated non-linear Pinney differential equation \cite{PINNEY} and rebuild explicitly the many-body wave function from which exact solutions for the physical observables and in particular for the particle density are available  \cite{EL3,MING}. 

The paper is organized as follows: in the next section we present the model and its mapping to a Fermi system in the limit of hard-core bosons.  After the explicit diagonalization of the instantaneous Hamiltonian which is performed in section \ref{AD} we present the results obtained for the density profile in the first order time-dependent perturbation theory framework in section \ref{QAD}. The dynamical invariants approach is presented in section \ref{EX} where the exact result for the density profile, obtained in the case of a linear ramp, is compared to the first order perturbative one. The release of the harmonic trap with a frequency that varies as the inverse of the time is also considered there and solved explicitly. It is shown in particular that the bosonic cloud expands with a power-law behaviour which exponent is a continuous function of the initial amplitude of the trap. At high initial amplitudes, the exponent becomes complex and leads to the appearance in the expansion of the cloud to a log-periodic modulation in time of square-root growing law.  
Finally, a brief summary and conclusions are given in the last section.

\section{The model}\label{MOD}

The Hamiltonian of the one dimensional Bose-Hubbard model, which describes a set of bosons living on a lattice with repulsive on-site interaction $U$ and submitted to an external time-dependent potential $V(t)$ is given by
\be
H= -\frac{J}{2} \sum_{j=-L/2}^{L/2} [a^\dagger_{j+1} a_j + h.c.] + \frac{U}{2}\sum_{j=-L/2}^{L/2} n_j (n_j-1) + \sum_{j=-L/2}^{L/2} V_j(t) n_j
\ee 
where the creation and annihilation operators $a^\dagger,a$ satisfy the usual canonical bosonic algebra $[a_j,a^{\dagger}_k]=\delta_{j,k}$, $[a_j,a_k]=[a^\dagger_j,a^\dagger_k]=0$ and where $n_j=a^\dagger_j a_j$ stands for the occupation number at site $j$. 
The first term proportional to $J$ describes the kinetic part of the system and in the following we will set the hopping amplitude $J=1$. 
In the hard core limit, that is for a very large repulsive interaction $U/J\gg 1$, the Bose-Hubbard Hamiltonian reduces to 
\be
H= -\frac{1}{2} \sum_{j=-L/2}^{L/2} [b^\dagger_{j+1} b_j + h.c.] + \sum_{j=-L/2}^{L/2} V_j(t) n_j
\label{TGH}
\ee  
with a new set of operators $b^\dagger, b$ that still satisfy the bosonic algebra for different sites but fulfills the on-site anti-commutation relations 
$\{b^\dagger_j,b_j\}=1$, $\{b_j,b_j\} = \{b^\dagger_j,b^\dagger_j\}=0 $, 
which prevent a double occupancy of a given site (the occupation operator is $n_j=b^\dagger_j b_j$). Obviously, the Pauli raising and lowering operators $\sigma_j^+$ and $\sigma_j^-$ realize the algebra generated by the operators $b_j^\dagger$ and $b_j$. 

The standard procedure to diagonalize the Tonks-Girardeau (TG) Hamiltonian (\ref{TGH}) is first to fermionize it through the Jordan-Wigner transformation \cite{JordanWigner}, mapping the operators $b$ with mixed bosonic and fermionic characters to simple fermionic $c$ operators:
\be
c_j^{\dagger}=\prod_{i<j}(1-2b^{\dagger}_ib_i)b^{\dagger}_j 
\label{JW}
\ee
and the associated mapping for the adjoint annihilation operators $c_j$. 
Under this transformation,  the TG Hamiltonian reduces to a spinless tight-binding Fermi system. Explicitly, for a finite size lattice with open boundary conditions one has 
\be\label{A}
H=\sum_{i,j=-L/2}^{L/2} c_i^{\dagger} A_{i,j}(t) c_j\,,
\ee
where we have introduced the matrix $A(t)$: 
\be
A_{i,j}(t)\equiv V_i(t)\delta_{i,j} -\frac{1}{2}(\delta_{i,j+1}+\delta_{i+1,j})\; .
\ee
Notice that the occupation number operator $n_j=b^{\dagger}_jb_j=c^{\dagger}_jc_j$. 
At a given time $t$ the quadratic form (\ref{A}) is readily diagonalized through a unitary transformation reducing $(\ref{A})$ into a free theory:
\be
H= \sum_{q=0}^L \epsilon_q(t) \eta^{\dagger}_q(t)\eta_q(t)\; ,
\ee
where the $\epsilon_q(t)$ are the single particle energies and  where the  diagonal Fermi operators $\eta^\dagger$, $\eta$ are defined through
\be
\eta^{\dagger}_q(t)=-\sum_{i=-L/2}^{L/2} \psi_q(i,t) c_i^{\dagger}\; 
\label{eta}
\ee
and the associated relation for the adjoints $\eta_q(t)$. Notice here that the time $t$ appears in these expressions as a simple parameter. 
The Bogoliubov coefficients $\psi_q(i,t)$ satisfy the  orthonormality condition $\sum_i \psi^*_q(i,t)\psi_p(i,t)=\delta_{qp}$ which in turn 
implies the canonical anti-commutation algebra $\{\eta^\dagger_q(t), \eta_p(t)\}=\delta_{qp}$,  
$\{\eta^\dagger_q(t), \eta^\dagger_p(t)\}=\{\eta_q(t), \eta_p(t)\}=0$.
The minus sign in (\ref{eta}) is irrelevant and set for further conveniences.\\
The time-dependent  potential $V(t)$ confining the Bose gas is 
\be\label{trap}
V_j(t)= |g(t)|j^2 -\mu
\ee
where the shift $\mu$, with $-1<\mu< 1$, can be interpreted as a chemical potential and the time-dependent amplitude $g(t)$ is assumed to be a slowly-varying function (this will be made precise later).

\section{Instantaneous diagonalization and adiabatic evolution}\label{AD}
\subsection{Instantaneous diagonalization}

\begin{figure}
\centering
\includegraphics[width=0.6\textwidth]{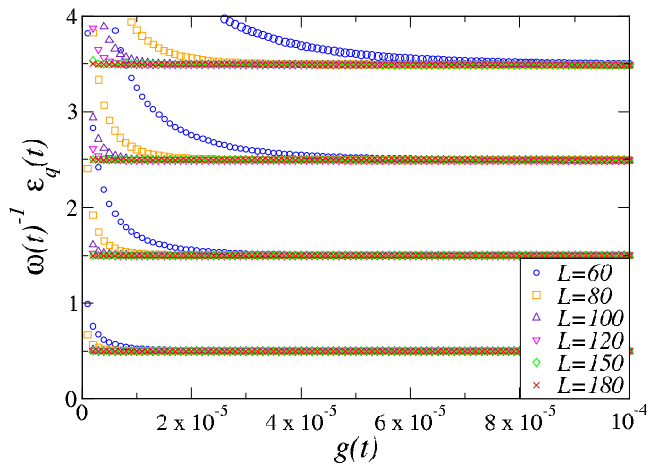}
\caption{
The  $q=0,1,2,3$ rescaled single particle energies  $\omega^{-1}(t)\epsilon_q(t)$ obtained from the exact diagonalization of the matrix $A(t)$ as a function of $g(t)$ and for different system sizes.}\label{energy}
\end{figure}

\begin{figure}
\centering
\includegraphics[width=0.9\textwidth]{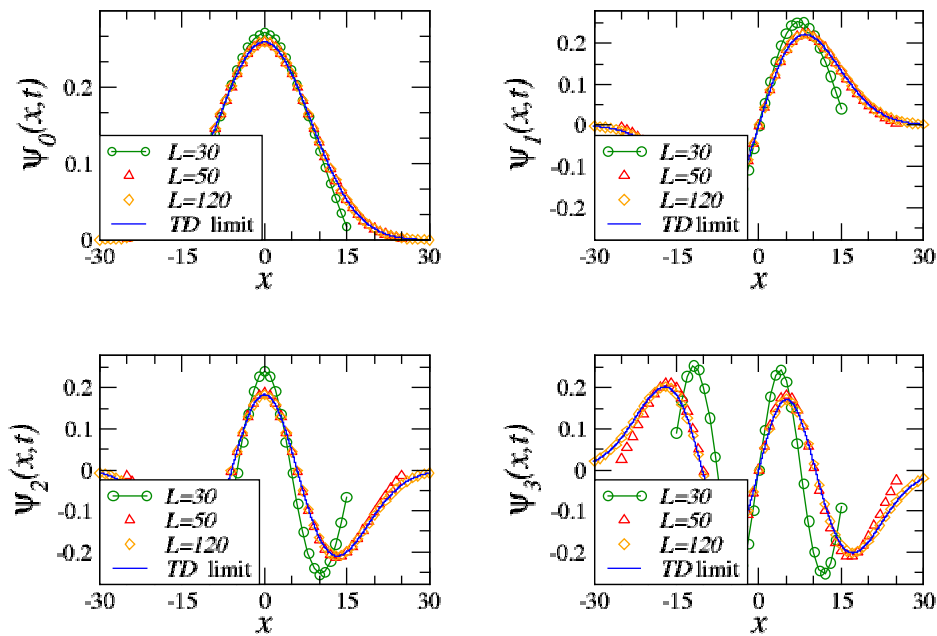}
\caption{
The  $q=0,1,2,3$ eigenvectors obtained from the exact diagonalization of the matrix $A(t)$ at $g(t)=10^{-4}$ for different system sizes compared to the thermodynamic limit.}\label{eigenvect}
\end{figure}

The instantaneous single particle energy spectrum and the corresponding eigenvectors are derived through the diagonalization of the matrix $A(t)$ (notice that here the time variable $t$ is just a parameter):
\be\label{eigendiscrete}
A(t)\psi_q(t)=\epsilon_q(t) \psi_q(t)\; .
\ee
Numerical exact diagonalization are easily performed for such a problem (see figure \ref{energy}). However, 
in the thermodynamic limit, in which the lattice site $i$ is replaced by a continuous variable $x=a i$, and expanding
\be
\psi_q(x_0\pm a,t) \simeq \psi_q(x_0,t) \pm a\, \de_x \psi_q(x_0,t) +\frac{a^2}{2!}\,\de^2_x \psi_q(x_0,t)\; ,
\ee
the eigenvalue problem $(\ref{eigendiscrete})$ reduces to
\be
\frac{1}{2}\de^2_x \psi_q(x,t)+ [\epsilon_q(t)+1-V(x,t)]\psi_q(x,t)=0 \; ,
\ee
where $V(x,t)=|g(t)| x^2-\mu$ is the continuum limit of the lattice potential of the Eq. $\eqref{trap}$. Choosing $\mu=-1$ (which corresponds in the absence of the trapping potential to the transition point  between the trivial Mott phase with zero density for $\mu<-1$ and the superfluid phase for $|\mu|<1$) one obtains the stationary Schr\"odinger equation for the $1d$ quantum harmonic oscillator
\be
\frac{1}{2}\Big(-\de_x^2+\omega^2(t)x^2\Big) \psi_q(x,t)=\epsilon_q(t)\psi_q(x,t)\; , \qquad \omega(t)=\sqrt{2|g(t)|}
\ee
with energies $\epsilon_q$ and corresponding eigenfunctions $\psi_q$.  Explicitly the solution is
\be
\label{psi}
\psi_{q}(x,t)= \sqrt{\frac{\sqrt{\omega(t)}}{2^q q! \sqrt{\pi}}} e^{-\frac{\omega(t)}{2}x^2} \; {\rm He}_q(x\sqrt{\omega(t)})\;, \quad 
\epsilon_{q}(t)= \omega(t)(q+\frac{1}{2})\;, 
\ee
where $q\in \mathbb{N}$ and ${\rm He}_q$ denotes the $q^{th}$ Hermite polynomial with physical normalization implied from the normalization of the single particle wave functions $\psi_{q}(x,t)$.  In figure \ref{energy} we show the exact low-lying single particle rescaled energies as a function of the trap amplitude $|g(t)|$ and compare them to the continuum limit prediction given above. As the system size is increased the agreement gets better and better. 
The associated lowest eigenvectors for different sizes  are shown in figure \ref{eigenvect}. 

The frequency $\omega(t)$ defines an instantaneous length scale in the problem which is given by $\ell(t)\equiv(\omega(t))^{-1/2}$. In terms of this length scale, the solution is 
\be
\psi_q(x,t)= \ell^{-1/2} \, \chi_q\left(\frac{x}{\ell}\right)\; ,
\qquad \chi_q(u)
= \frac{1}{\sqrt{2^q q! \sqrt{\pi}}} e^{-\frac{1}{2}u^2} \; {\rm He}_q(u) 
\label{psiscaling}
\ee
for the eigenstates, and
\be
\label{escaling}
\epsilon_q(t)= \ell^{-2} (q+\frac{1}{2})
\ee
for the energy spectrum, where the time-dependence of $\ell(t)$ has been implicitly considered. This scaling form for the eigenfunctions and energies is in agreement with  general time-dependent trap-size scaling  arguments. Indeed, 
for a driven inhomogeneous quantum quench through the MI-SF critical point, the inhomogeneous control parameter 
\be
\delta\mu(x,t)\equiv\mu(x,t)-\mu_c \approx-|g(t)|x^w 
\ee
induces a finite-length scale $\ell(t)$ \cite{gradient,PKT07,critical,nonlin,VIC3}. This typical length scale can be derived self-consistently from the space dependence of the quantum deviation parameter by
\be
\ell(t)\propto |\delta\mu(\ell(t),t)|^{-\nu} 
\ee
where $\nu$ is the correlation critical exponent. The solution of this equation is 
\be
\label{ell}
\ell(t) \propto |g(t)|^{-\nu_g}\; , \qquad \nu_g=\frac{\nu}{1+\nu w}
\ee
The scaling behavior of a local quantity $\varphi(x,t)$ with scaling dimension $x_{\varphi}$ is expected to be
\be
\varphi(x,t) \propto \ell^{-x_{\varphi}} \, \tilde{\Phi}(\frac{x}{\ell})\;,
\ee
where $\tilde{\Phi}$ is a scaling function. At the MI-SF transition, where the critical exponents are $\nu=1/2$ for the correlation length and  $z=2$ for the dynamics, fixing $w=2$ for a parabolic trap one recovers the scaling forms (\ref{psiscaling}) and (\ref{escaling}) with $\nu_g=1/4$ and $x_\psi=1/2$ for the scaling dimension of the single particle wave function.
Notice that for a finite size system, the size $L$ of the system itself becomes a scaling field and one expects that a given quantity depends on the two length scales $\ell(t)$ and $L$, such that
\be
\varphi(x,L,t) \propto \ell^{-x_{\varphi}} \, \tilde{\Phi}(\frac{x}{\ell}, \frac{L}{\ell}) \; . 
\label{scalL}
\ee
For $L\ll \ell(t)$ it is expected that the scaling relation $(\ref{scalL})$ matches the ordinary finite-size scaling behavior $\varphi  \propto L^{-x_\varphi}$ while for $L\gg \ell(t)$ the system becomes independent on the lattice size $L$ and matches the infinite volume behaviour. The thermodynamic limit is therefore taken as the limit $L\To\infty$, $\ell \To \infty$ with the ratio $L/\ell^2$ (or equivalently $L^2 g$) fixed such that finite size corrections are avoided. 

\subsection{Adiabatic evolution of the density}
\begin{figure}
\centering
\includegraphics[width=0.9\textwidth]{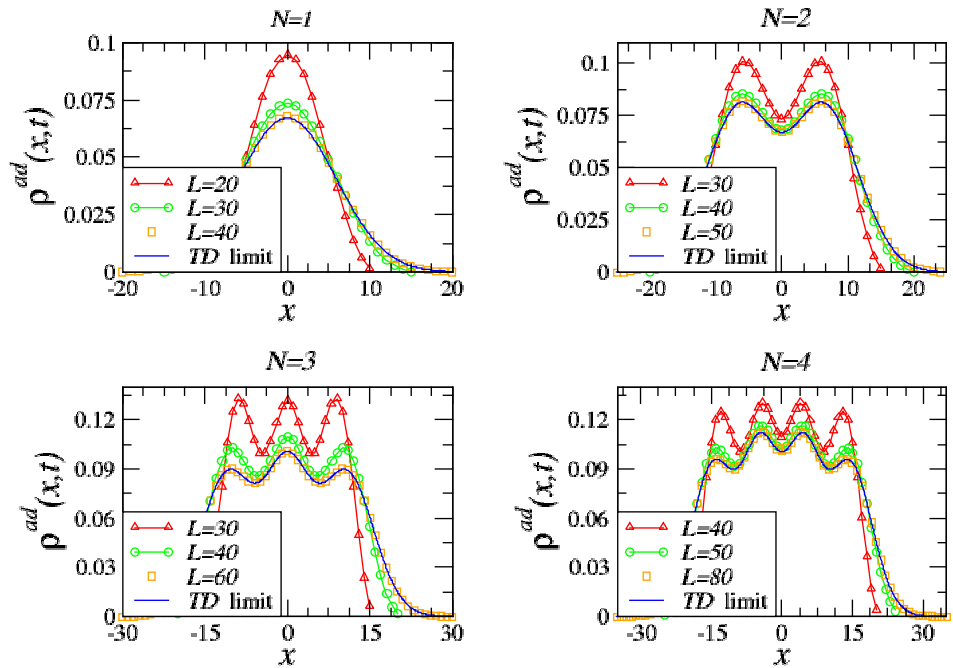}
\caption{The density profile of the system for an adiabatic evolution for a different number $N$ of particles at a time $g(t)=10^{-4}$. Numerical results for different lattice sizes are compared with the analytical results in the thermodynamic limit. }\label{densAD}
\end{figure}

\begin{figure}
\centering
\includegraphics[width=0.9\textwidth]{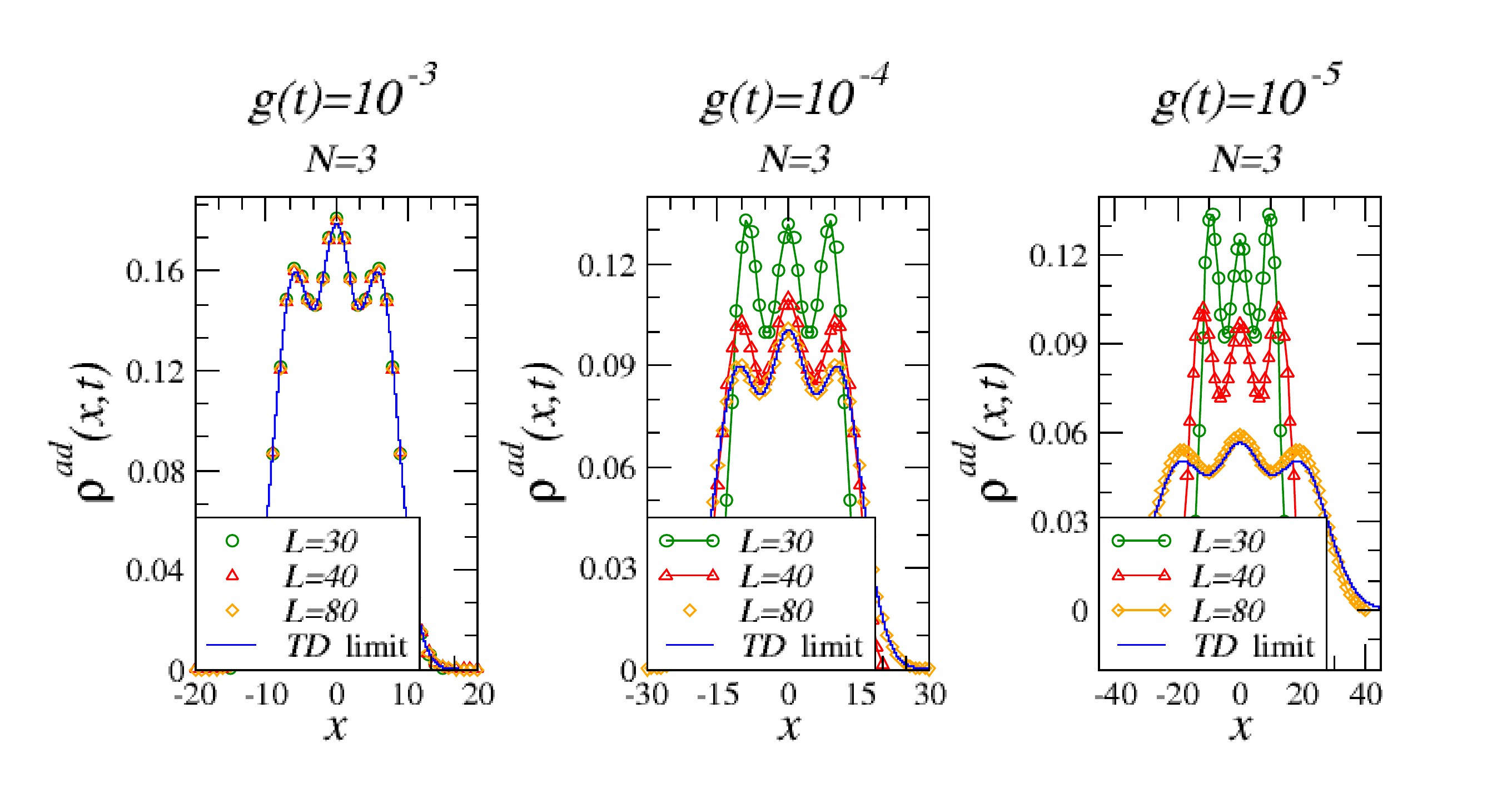}
\caption{The density profile of the system for an adiabatic evolution for $N=3$ particles for three different values of the trap amplitude $g(t)=10^{-3},\; 10^{-4},\; 10^{-5}$.   Numerical results for different lattice sizes are compared with the analytical results in the thermodynamic limit. The bigger the amplitude is the smaller the system size needs to be in order to achieve the thermodynamic limit. }\label{densAD2}
\end{figure}

\begin{figure}
\centering
\includegraphics[width=0.9\textwidth]{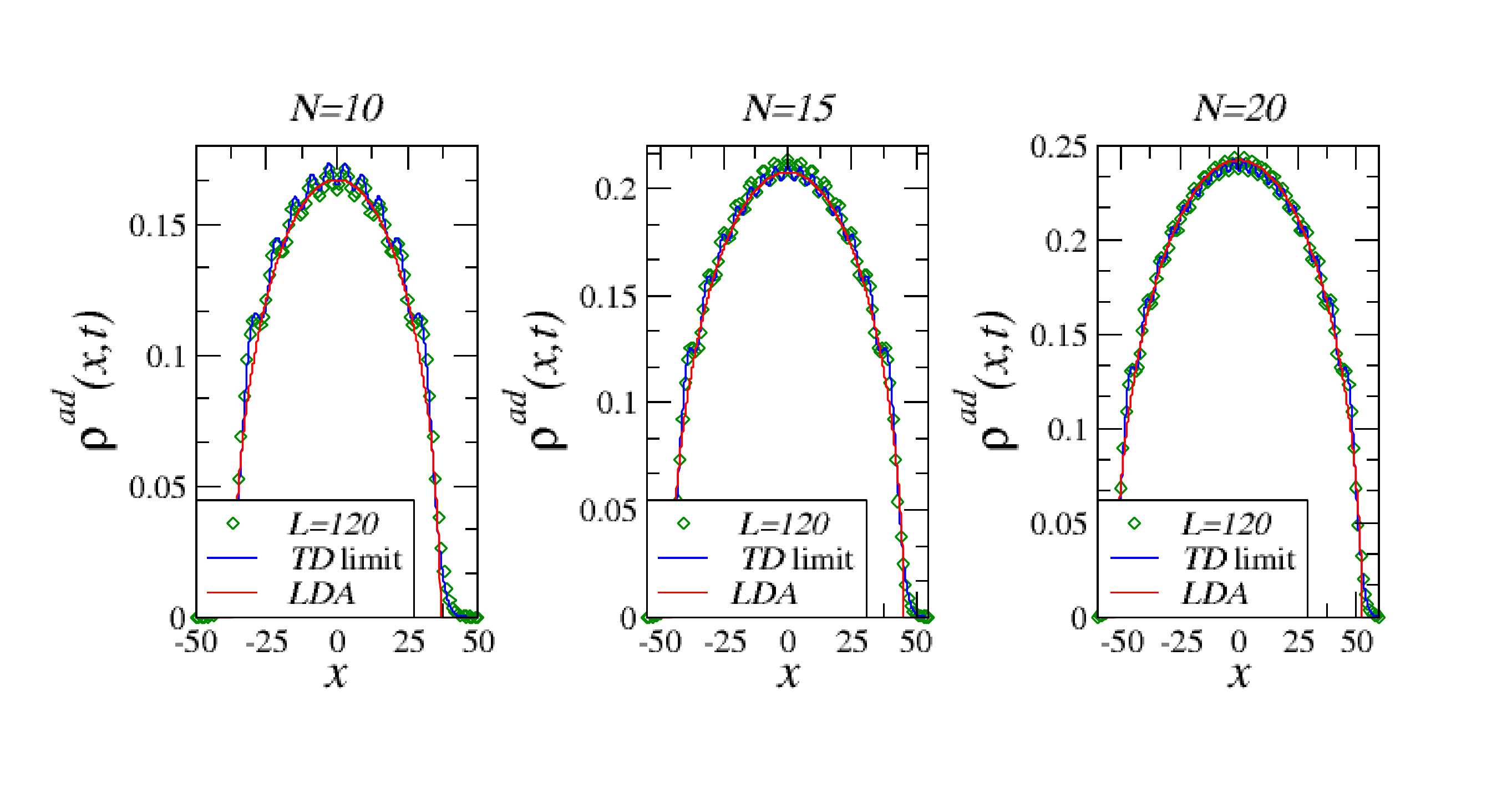}
\caption{The adiabatic density profile at $g(t)=10^{-4}$ for a large number $N$ of bosons. The plots show the numerical results and the thermodynamic analytical results compared with the local density approximation value.}\label{LDA}
\end{figure}

At the initial time $t_0$ the system is prepared in the $N$-particles ground state of (\ref{TGH}) which is given by
\be
\label{GS0}
|\s_0(t_0)\rangle = \prod_{q=0}^{N-1} \eta^{\dagger}_q(t_0)| 0\rangle \; , 
\ee
where $|0\rangle$ is the vacuum state such that $\eta_q(t_0)| 0\rangle = 0$ $\forall q$, since all the single particle energies are positive.
The energy associated to the initial state $|\s(t_0)\rangle$ is thus simply given by
\be
E_0(t_0) = \sum_{q=0}^{N-1} \epsilon_q(t_0)\; .
\ee
For a very slow variation of the confining potential, it is expected that the system adapts itself to the instantaneous Hamiltonian $(\ref{TGH})$ and evolves remaining in the instantaneous ground state $|\s_0(t)\rangle = \prod_{q=0}^{N-1} \eta^{\dagger}_q(t)| 0\rangle$. 
The evolution of the particle density is thus expected to be given by the adiabatic density
\begin{eqnarray}
\label{rhoad}
\rho^{ad}(i,t)&=& \langle \s_0(t)|c^{\dagger}_ic_i|\s_0(t)\rangle \nonumber \\
&=& \prod_{k,k'=0}^{N-1}  \sum_{q,q'=0}^L\, \psi^{\ast}_q(i,t)\psi_{q'}(i,t)\; \langle 0| \eta_{k}(t) \eta^{\dagger}_q(t)\eta_{q'}(t)\eta^{\dagger}_{k'}(t)|0\rangle \nonumber \\
&=& \sum_{k=0}^{N-1} |\psi_k(i,t)|^2 \; ,
\end{eqnarray}
or in the thermodynamic limit by the scaling form 
\be\label{rhoAD}
\rho^{ad}(x,t) = \ell^{-1} \, f^{ad}\left(\frac{x}{\ell}\right)\;, \qquad  f^{ad}(u)=\sum_{q=0}^{N-1} |\chi_q(u)|^2\; ,
\ee
where the functions $\chi_q$ are defined in (\ref{psiscaling}). 
In figure \ref{densAD} we show the convergence of the adiabatic density profile obtained from exact numerical diagonalization toward the thermodynamical limit expression for small particle numbers $N$. 
In the large size limit the exact numerical results match perfectly the analytical expression (\ref{rhoAD}). At smaller sizes the finite size corrections to the excitation spectrum and to the corresponding eigenvectors lead to quite a discrepancy between the thermodynamic limit result and the actual finite size density profile.
In figure \ref{densAD2} we show the adiabatic density for $N=3$ bosons for three different values of the amplitude $g(t)$. 
\\
For a large number of bosons the adiabatic density profile matches its Local Density Approximation (LDA) limit. The LDA is obtained assuming that for each instant of time $t$ and around each (coarse-grained) point $x$ there is a local flat band of excitations with dispersion $\varepsilon(x)=-\cos q_F(x)+V(x,t)$. Locally the single particle band is filled up to the global Fermi level given here by $\varepsilon_F=\varepsilon_{N-1}(t)$ and the LDA $\rho^{LDA}(x)$ is deduced from the associated local Fermi momentum $q_F(x)=\pi \rho^{LDA}(x)$ leading to \cite{CARK, VIC, Pierre}
\be
\rho^{LDA}(x,t)=\frac{1}{\pi}\, \arccos(V(x,t)-\epsilon_{N-1}(t))\;.
\label{LDA1}
\ee
This is shown in figure \ref{LDA} for three different values of $N$. The exact numerical profile, as obtained from exact diagonalization, is compared to the thermodynamic limit (\ref{rhoAD}) and LDA (\ref{LDA1}) results. 
One can recover the scaling form (\ref{rhoAD}) from (\ref{LDA1}) in the limit $N\ll \ell^2$. Indeed, using 
$\ell=(\omega(t))^{-1/2}=(2|g(t)|)^{-1/4} $ and the expression of the potential $V(x,t)=1 + |g(t)| x^2$ one has 
\be
\rho^{LDA}(x,t)=\frac{1}{\pi}\, \arccos\left( 1- \frac{1}{2\ell^2} \left[ 2N-1 -\left(\frac{x}{\ell}\right)^2   \right]\right) \; .
\ee
For $N/\ell^2 \ll 1$, expanding the $\arccos$ function to the leading order in $N^{1/2}$, one obtains a semi-circle law
\be
\label{semicircle}
\rho^{LDA}(x,t)\simeq \frac{1}{\pi \ell} \left( 2N-1 -\left(\frac{x}{\ell}\right)^2 \right)^{1/2} \;  \underset{N \gg 1}{\simeq} \;
\frac{\sqrt{2N}}{\pi \ell} \left(1-\frac{1}{2} \left(\frac{x}{\ell \sqrt{N}}\right)^2\right)^{1/2}
\ee
in agreement with the scaling form (\ref{rhoAD}). The support of the adiabatic density profile is in $[-\ell_N,\ell_N]$ with a number of particles typical length scale
\be
\ell_N = \ell \sqrt{2N-1} \underset{N \gg 1}{\simeq}  \ell\sqrt{2N} \; .
\ee 
In terms of that typical length scale $\ell_N$, the adiabatic profil takes the scaling form
\be
\label{semicircle2}
\rho^{LDA}(x,t)\simeq \frac{N}{\ell_N}\;  f^{LDA}\left(\frac{x}{\ell_N}\right)\; , \quad f^{LDA}(u)= \frac{2}{\pi} \sqrt{1-u^2}\; .
\ee

\section{First order correction to the adiabatic evolution}\label{QAD}
\subsection{First order correction to adiabaticity}
The first order correction to the adiabatic evolution, starting from an initial $N$ particles ground state $\ket{\s_0(t_0)}$ as defined in (\ref{GS0}), is given by \footnote{see the Appendix A of the work in Ref. \cite{nonlin}}:
\be\label{onejump}
 \ket{\s(t)}\simeq e^{-i\int_{t_0}^t dt'\, E_0(t')} \Big[ \ket{\s_0(t)} + \sum_{p=N}^L \sum_{k=0}^{N-1} a_{p,k}(t)\,  \ket{\s_0[\slashed{k},p\,] (t)}\Big]
 \ee
where $\ket{\s_0[\slashed{k},p \,] (t)}\equiv \eta^{\dagger}_p(t)\, \eta_k(t) \ket{\s_0(t)}$ is the instantaneous ground state in which a particle has been promoted from the lowest levels $k=0,\dots,N-1$ ($\slashed{k}$ denotes a vacancy in the position $k$) toward higher ones  $p=N,\dots,L$ (since the particles are fermionic in nature double occupancy of a state is forbiden). The transition amplitude $a_{p,k}(t)$  is given by
\be
\label{apert}
a_{p,k}(t)= \int_{t_0}^t dt' \, \frac{\de_{t'}K_{p,k}(t')}{\epsilon_p(t')-\epsilon_k(t')}\, \exp[-i\int_{t'}^t dt''\, (\epsilon_p(t'')-\epsilon_k(t''))]\; ,
\ee
where 
\be
 \qquad K_{p,k}(t) \equiv \sum_{j=-L/2}^{L/2}\psi_p^{\ast}(j,t) \,  V_j(t) \, \psi_k(j,t) 
\ee
is the instantaneous transition amplitude of the perturbation
\be
\delta H(t)\equiv \sum_{j=-L/2}^{L/2} V_j(t)\, n_j= \sum_{p,k=0}^{L} K_{p,k}(t)\, \eta^{\dagger}_p(t)\, \eta_k(t)\; .
\ee

In the scaling limit $g\To 0$, $L\To \infty$, the first order transition amplitude $a_{p,k}(t)$ can be computed analytically 
using the thermodynamic limit expressions (\ref{psiscaling}) and (\ref{escaling}).  After a straightforward computation one obtains 
\be\label{a}
a_{p,k}(t) = C_{p,k} \, \delta_{p-2,k}\, \ln\left(\frac{\ell(t)}{\ell(t_0)}\right)\; , \quad C_{p,k}= \frac{\sqrt{p(k+1)}}{ k-p}\; .
\ee 
Starting from the initial $N$ particles ground state $\ket{\s_0(t_0)}$, as seen from (\ref{onejump}) at the first order in perturbation only  
the lowest energy levels are activated: $(p,k)=(N+1,N-1)$ and for $N>1$, $(p,k)=(N,N-2)$. A crude approximation $p\simeq k\simeq N\gg 1$  shows that the transition amplitude $(\ref{a})$ is of order $N\, \ln \frac{\ell(t)}{\ell(t_0)}$ . Consequently, the order of the approximation (\ref{onejump}) is 
\be
\label{epsilong}
\varepsilon ={\cal O}\left( N \ln \frac{\ell(t)}{\ell(t_0)}\right) =  {\cal O}\left(N \nu_g \ln \frac{g(t)}{g(t_0)} \right) = {\cal O}\left(N \nu_g \frac{\delta g(t)}{g(t_0)}\right) \; ,
\ee
with $\delta g(t)\equiv g(t)-g(t_0)$.

\subsection{Particle density}
\begin{figure}
\centering
\includegraphics[width=0.9\textwidth]{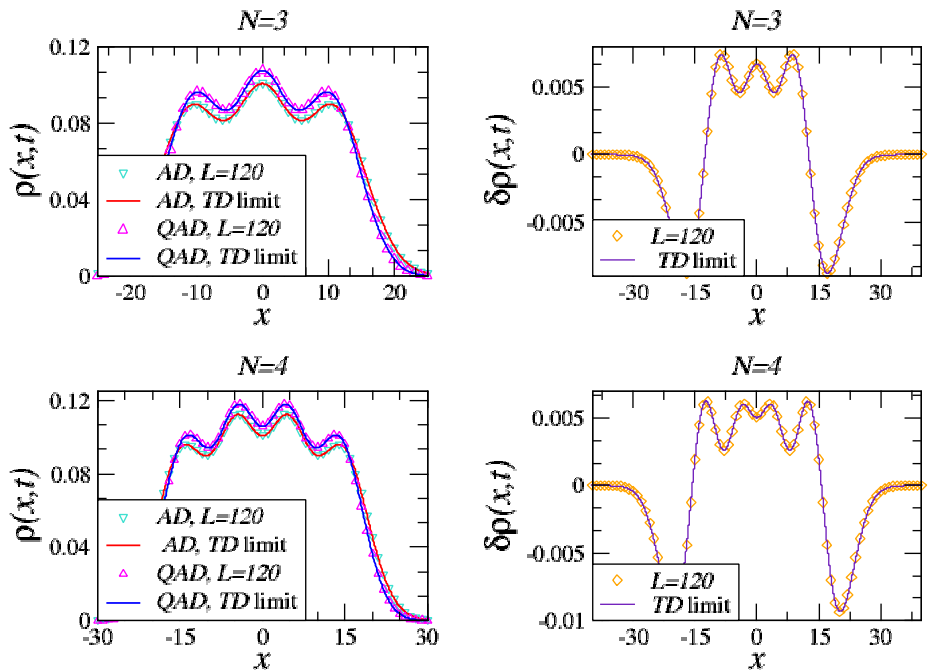}
\caption{\textsl{Left.} The density profile at first-order in perturbation theory compared with the adiabatic evolution for $N=3,4$ . \textsl{Right.} The first-order corrections to the adiabatic density profile for  $N=3,4$. The figures are made with $g_0=10^{-4}$ and fixing the precision of the expansion $\varepsilon=0.2$.}\label{quasiad}
\end{figure}

\begin{figure}
\centering
\includegraphics[width=0.9\textwidth]{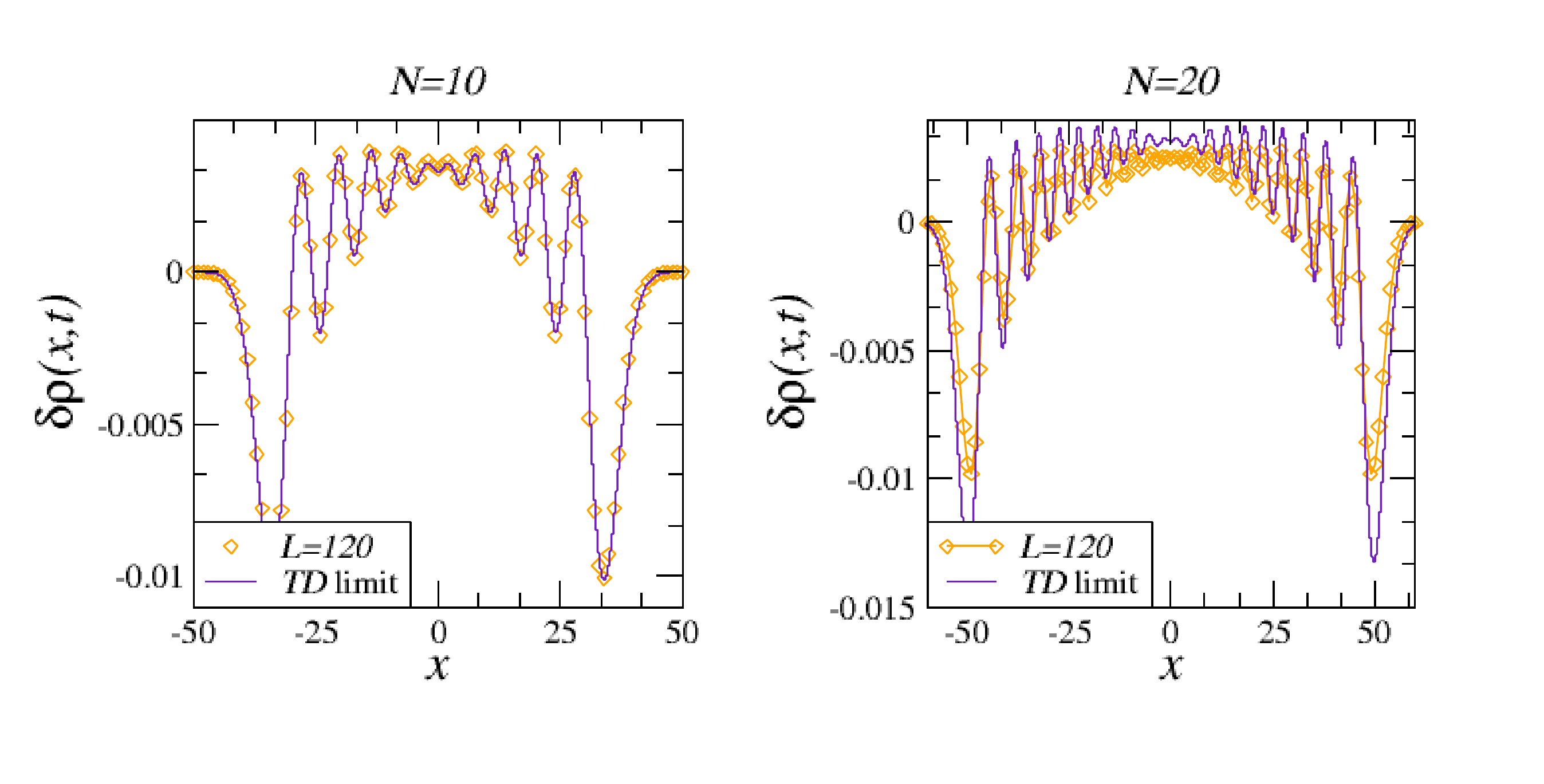}
\caption{The first-order corrections to the density profile for a large number $N$ of particles. The plot have been made for $\varepsilon=0.2$ and $g_0=10^{-4}$ .}\label{largeqad}
\end{figure}

\begin{figure}
\centering
\includegraphics[width=0.9\textwidth]{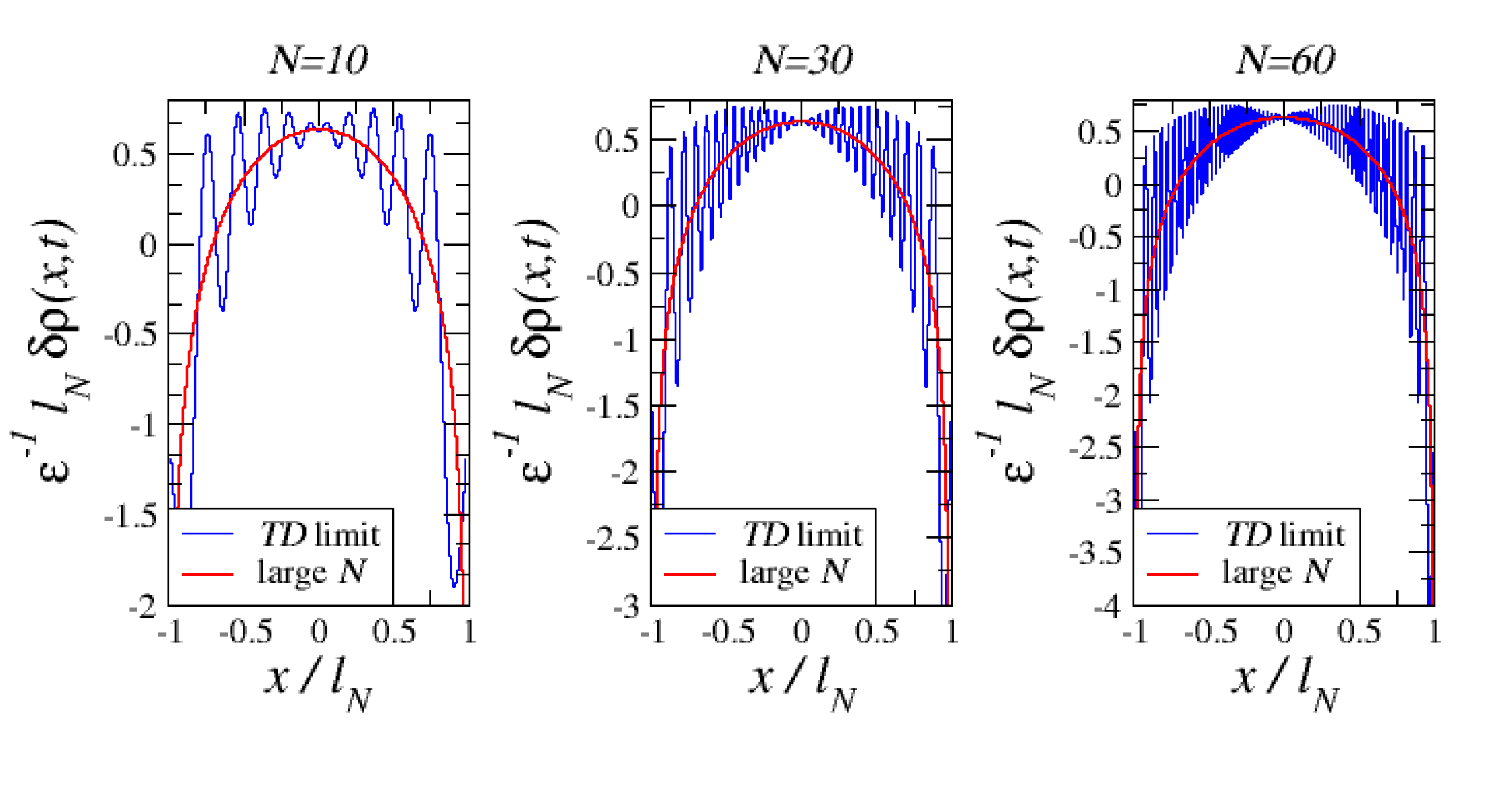}
\caption{The asymptotic behavior of the first-order corrections to the density profile $\eqref{Drho}$ for a different number $N$ of bosons compared with the result $\eqref{QADresult}$.}\label{f}
\end{figure}

The particle density at site $i$ and time $t$  is given by 
\be
\rho(i,t)=\sum_{q,q'=0}^L  \, \psi^{\ast}_q(i,t) \,w_{q,q'}(t)\,  \psi_{q'}(i,t) \; ,
\ee
where the two-point function 
\be
w_{q,q'}(t) \equiv \braket{\s(t)|\eta^{\dagger}_q(t)\eta_{q'}(t)|\s(t)}\; .
\ee
With the expansion (\ref{a}) at the leading order in $\varepsilon$ the two-point function is  given by
\bea
w_{q,q'}(t) = \sum_{k=0}^{N-1} \delta_{q,k}\, \delta_{q',k} &+& \sum_{p=N}^L\sum_{k=0}^{N-1} a^{\ast}_{p,k}(t)\,  \delta_{q',k} \, \delta_{q,p} \nonumber \\
&&+ \sum_{p'=N}^L\sum_{k'=0}^{N-1} a_{p',k'}(t) \, \delta_{q',p'} \, \delta_{q,k'} +{\cal O}(\varepsilon^2)\; .
\eea
The first term of the two-point function gives the adiabatic contribution (\ref{rhoad}). The deviation to the adiabatic density is thus expressed as
\be
\delta \rho(i,t)\equiv \rho(i,t)- \rho^{ad}(i,t) = 
\left(\sum_{q=N}^L \sum_{q'=0}^{N-1} a_{q,q'}^{\ast}(t)\, \psi_q^{\ast}(i,t)\, \psi_{q'}(i,t) + {\rm c. c.}\right) +{\cal O}(\varepsilon^2)\; .
\ee
In the scaling limit $g\To 0$, $L\To \infty$, using (\ref{psiscaling}) and  (\ref{a}) we obtain
 \bea
 \label{QADrho}
 \delta\rho(x,t)= \frac{2}{\ell(t)} \ln \frac{\ell(t)}{\ell(t_0)} &&\left[
 C_{N,N-2}\; \chi_{N}\left(\frac{x}{\ell(t)}\right) \, \chi_{N-2}\left(\frac{x}{\ell(t)}\right)\right. \nonumber \\
 &&+ \left. C_{N+1,N-1}\; \chi_{N+1}\left(\frac{x}{\ell(t)}\right) \, \chi_{N-1}\left(\frac{x}{\ell(t)}\right)�\right]
 \eea
with $C_{1,-1}=0$ by convention. Using recursion relations of the Hermite polynomials the density deviation $\delta\rho(x,t)$
can be rewritten as 
\be\label{QADresult}
\delta\rho(x,t)= \frac{1}{2\ell(t)} \ln \frac{\ell(t)}{\ell(t_0)} \left[ F_N\left( \frac{x}{\ell(t)}\right) +  F_{N-1}\left( \frac{x}{\ell(t)}\right) \right]
\ee
where $F_N(u)$ is given by
\be
\label{FNu}
F_N(u) =   (\chi_N'(u))^2  - u^2 \chi_N^2(u) \; .
\ee
A plot of the density profile in the quasi-adiabatic evolution for low values of $N$ is shown in figure \ref{quasiad}.\\
At large particle number $N$, the Hilbert-Hermite functions $\chi_N(u)$ take significant values only in the region $|u|<\sqrt{2N}$ 
where the zeros of the Hermite polynomials are located. Outside that region the Hilbert-Hermite functions decay exponentially fast.
In the limit $N\gg 1$, it has been shown in \cite{Dominici} that the Hermite polynomials have the asymptotic representation for $\theta\in]-\pi/2,\pi/2[$ given by
\be
{\rm He}_N(\sqrt{2N} \sin \theta)\sim \left(\frac{2N}{e}\right)^{\frac{N}{2}}\sqrt{\frac{2}{\cos \theta}}e^{N\sin^2\theta}\; \cos h_N(\theta)
\ee
with the phase
\be
h_N(\theta)= N \left[\frac{1}{2}\sin(2\theta)+\theta -\frac{\pi}{2}\right] + \frac{\theta}{2}\; .
\ee
With Stirling formula $N!\simeq \sqrt{2\pi N} (\frac{N}{e})^N$ and the asymptotic representation given above, the Hilbert-Hermite functions,  in the limit $N\gg 1$, take the form
\be
\label{chiexp}
\chi_N(\sqrt{2N} \sin \theta)\sim  \frac{1}{(2N)^{1/4}}  \sqrt{\frac{2}{\pi \cos \theta}}\; \cos h_N(\theta)\; . 
\ee
Using this, one has for (\ref{FNu}) at large $N$ the asymptotic expression 
\be
F_N(u) 
\sim - \frac{2}{\pi} (2N)^{1/2} \frac{1}{\sqrt{1-\frac{u^2}{2N}}} \left[ \frac{u^2}{2N} -\sin^2h_N(u)\right]\; .
\ee
The $\sin^2h_N(u)$ gives a widely oscillating term and taking its average, $\sin^2h_N(u)\sim 1/2$, one finally obtains a scaling form for the deviation  $\delta\rho(x,t)$ as a function of the scaling variable $x/\ell_N$:
\be
\label{Drho}
\delta\rho(x,t)\sim  \frac{\varepsilon}{\ell_N} f^{\delta\rho}\left(\frac{x}{\ell_N}\right) \;, 
\quad f^{\delta\rho}(u)= \frac{4}{\pi}\; \frac{1/2-u^2}{\sqrt{1-u^2}}\; ,
\ee
where we have set the small parameter $\varepsilon$ associated to the first order correction  to 
\be
\label{varepsilon}
\varepsilon = N \ln \frac{\ell_N(t)}{\ell_N(t_0)} \; .
\ee
This behavior is shown in figure \ref{largeqad} and figure \ref{f} for the associated scaling function at large number $N$. \\
Notice that using the asymptotic expression (\ref{chiexp}) one has after averaging the $\cos^2$ term
\be
\chi_{q\gg 1}^2(u) \sim \frac{1}{\pi} \frac{1}{\sqrt{2q-u^2}} \theta(q-\frac{u^2}{2})
\ee
where $\theta(q)$ is the Heaviside function. The adiabatic density is thus 
\be
\rho^{ad}(x,t)= \frac{1}{\ell(t)} \sum_{q=0}^{N-1}\chi_{q}^2(u) \sim \frac{1}{\pi \ell(t)} \int_{u^2/2}^N \frac{{\rm d }q}{\sqrt{2q-u^2}}\sim
\frac{N}{\ell_N(t)} \frac{2}{\pi} \sqrt{1-\frac{u^2}{2N}}
\ee
which is nothing but the LDA semi-circle law (\ref{semicircle2}). 

\section{Exact Ermakov-Lewis evolution}\label{EX}
\subsection{Dynamical invariant approach}
\begin{figure}
\centering
\includegraphics[width=0.6\textwidth]{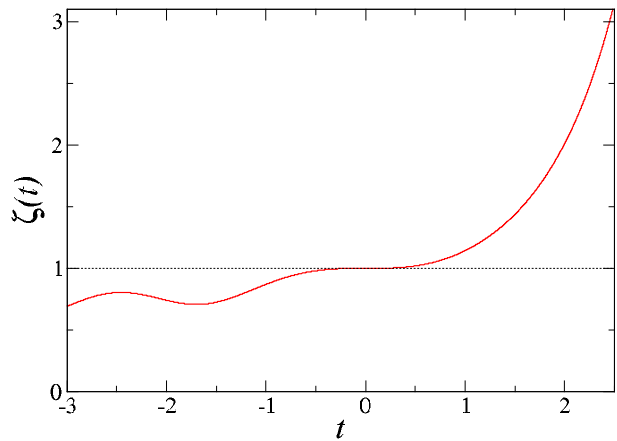}
\caption{The solution of the Eq.$\eqref{Pinney}$ for a linear ramp protocol $\eqref{g}$ for $\omega_0=1$ and $\alpha=1$.}\label{zeta}
\end{figure}

\begin{figure}
\centering
\includegraphics[width=0.9\textwidth]{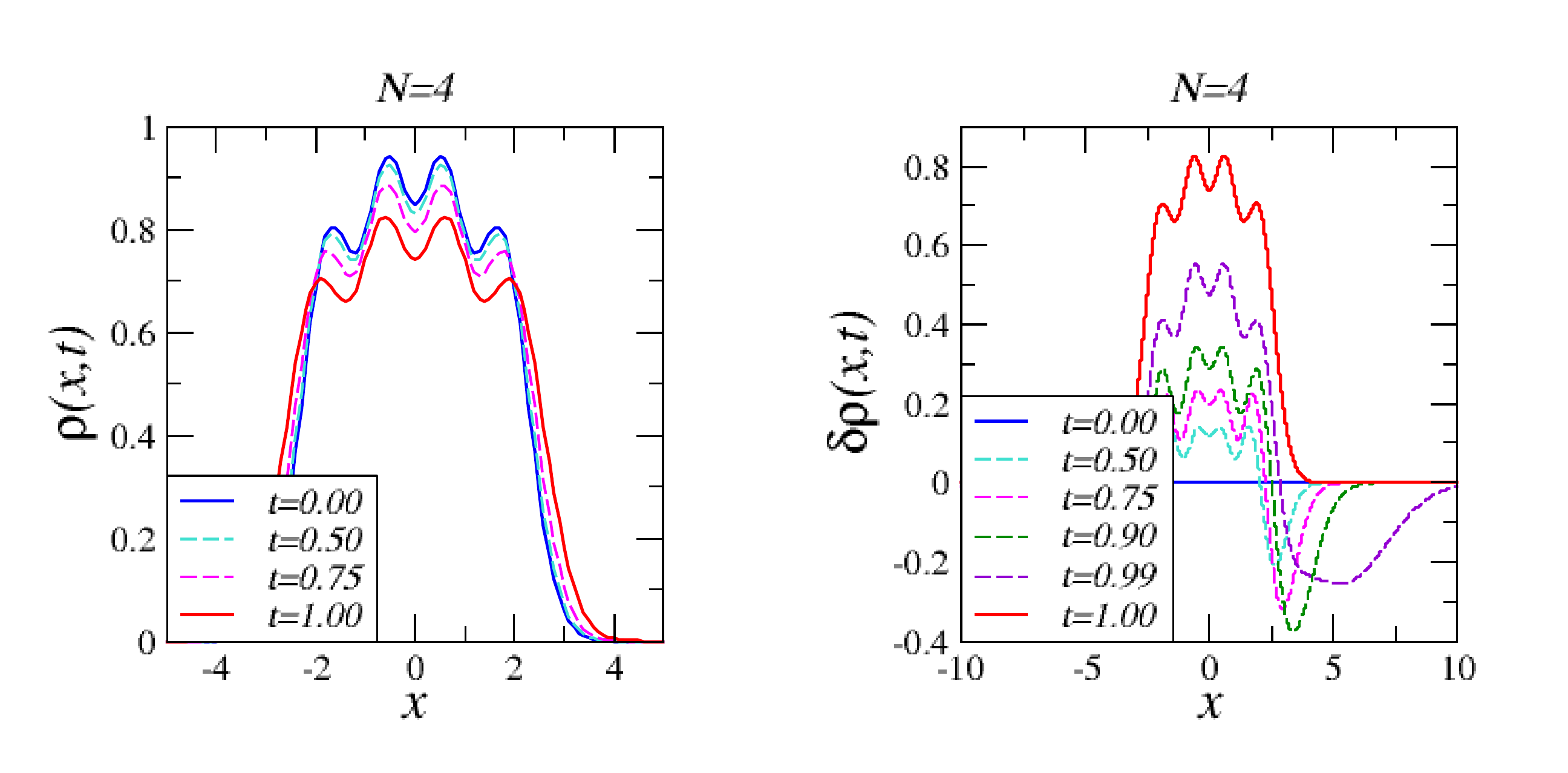}
\caption{\textsl{Left.} The exact off-equilibrium evolution of the cloud of $N=4$ bosons for a linear ramp protocol $\eqref{g}$. \textsl{Right.} The exact departure from the adiabaticity during the time-evolution of the cloud. For $t>1$ the cloud freerly expands.}\label{exact}
\end{figure}

The Ermakov-Lewis approach \cite{EL1,EL2} based on the identification of dynamical invariants is an alternative way that leads to exact results for particular ramps $g(t)$ .  Let us first reconsider the single-particle Schr\"odinger equation: 
\be
i\de_t \, \phi_q(x,t)= \frac{1}{2}(-\de_x^2 +\omega^2(t) x^2)\, \phi_q(x,t) \ee
with $\omega(t)=\sqrt{2|g(t)|}$ and suppose that the initial condition at $t_0$ is an eigenstate $\phi_q(x,t_0)=\psi_{q}(x,t_0)$ (see (\ref{psi}))
of the harmonic oscillator with initial pulsation  $\omega(t_0)\equiv \omega_0$. 
According to the dynamical invariant approach, the  time-evolved single-particle wave function $\phi_q(x,t)$ can be expressed as \cite{EL1,EL2,PINNEY,EL3}
\be\label{one}
\phi_q(x,t)= \frac{1}{\sqrt{\zeta(t)}} \, \exp\Big[i\frac{\DE{\zeta}(t) x^2}{2\zeta(t)}-i\omega_0(q+\frac{1}{2})\int_0^t \frac{dt'}{\zeta^{2}(t')}  \Big] \psi_q(\frac{x}{\zeta(t)},t_0)\; ,
\ee
where $\zeta(t)$ is the solution of the non-linear differential equation
\be\label{Pinney}
\DEE{\zeta}(t) +\omega^2(t)\zeta(t) = \omega^2_0 \zeta^{-3}(t) 
\ee
with initial conditions $\zeta(t_0)=1$ and $\DE{\zeta}(t_0)=0$.  The problem is thus reduced to solving this differential equation given the time-dependent protocol $g(t)$. For instance, setting a linear ramp
\be\label{g}
g(t)=g_0(1-  \alpha t)\; \qquad  t \leq 1, \; t_0=0
\ee
with time-scale $1/\alpha$, a solution for $\zeta$ is explicitly known in terms of Airy functions, see for example \cite{VIC,Pol1}, and is plotted in figure  \ref{zeta} as an illustration. The time-evolution of the density profile for the linear ramp protocol $\eqref{g}$ is shown in figure \ref{exact}.
%

From the knowledge of the single-particle wave functions $\eqref{one}$ we can write down the $N$-particles state as
\be
\Phi_N(\vec{x},t)= \frac{1}{\sqrt{N!}} \, \frac{\Delta(\vec{x})}{|\Delta(\vec{x})|} \, \det_{j,k=0}^{N-1} (\phi_k(x_j,t))
\ee
with $\vec{x}\equiv (x_0,x_1,...,x_{N-1})$. The Vandermonde determinant $\Delta$ symmetrizes the Slater determinant under particle exchange giving us the wave function of $N$ hard-core bosons since
\be
\frac{\Delta(\vec{x})}{|\Delta(\vec{x})|} = \prod_{i<j} \, \text{sgn}(x_i-x_j) \; .
\ee

Introducing the generating functional 
\be
\mathcal{Z}[a]= \frac{1}{N!} \, \int {\rm d}\vec{x} \; \prod_{j=1}^N a(x_j)  \, \det_{j,k=0}^{N-1}(\phi_k^{\ast}(x_j,t))\,  \det_{j,k=0}^{N-1} (\phi_k(x_j,t)) \; ,
\ee
the time-dependent particle density can be expressed as a functional derivative of $\mathcal{Z}[a]$:
\be
\rho(x,t)=\frac{\delta}{\delta a(x)}\Big\vert_{a\equiv 1} \mathcal{Z}[a(x)] = \int {\rm d} \vec{x} \; \Phi_N^{\ast}(\vec{x},t)\, \Phi_N(\vec{x},t) \, \sum_{j=0}^{N-1} \delta(x_j-x)\; .
\ee
Using the random matrix approach \cite{Rand,FORR}, we end up with the explicit result \cite{MING,VIC}
\be\label{rhoEX}
\rho(x,t)= \frac{1}{\zeta(t)} \sum_{k=0}^{N-1} \left|\psi_k\left(\frac{x}{\zeta(t)},t_0\right)\right|^2
= \frac{1}{\zeta(t)\ell_0} \sum_{k=0}^{N-1} \left|\chi_k\left(\frac{x}{\zeta(t)\ell_0},t_0\right)\right|^2
\ee
where $\ell_0\equiv\ell(t_0)$. The effect of the dynamics is completely absorbed in the definition of a non-trivial length scale 
\be
\label{xi}
\xi(t)=\ell(t_0) \, \zeta(t)\; .
\ee
The adiabatic limit is recovered by $\lim_{t\rightarrow t_0}\zeta(t)=1$, for which $\xi=\ell_0$. 
\subsection{Comparison with the quasi-adiabatic case}
\begin{figure}
\centering
\includegraphics[width=0.9\textwidth]{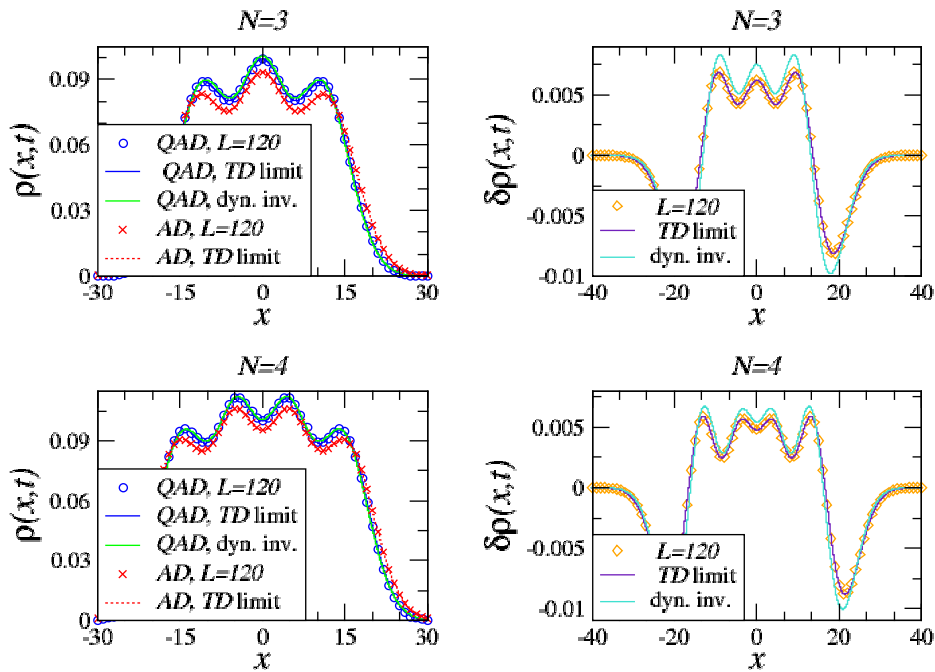}
\caption{\textsl{Left.} The density profile in the quasi-adiabatic limit for $N=3,4$. \textsl{Right.} Deviation from adiabaticity during the trap release for $N=3,4$. Numerical exact diagonalization, analytical results $\eqref{QADrho}$ in the thermodynamic limit and the density profile built with the use of dynamical invariants $\eqref{rhoEX}$ are compared. The figures are made fixing $\varepsilon=0.2$ and $t=4/N$.}\label{COMP}
\end{figure}

Let us consider the protocol (\ref{g}) in the quasi-adiabatic regime, i.e. when the quench rate $\alpha\To 0$. 
Expanding $\zeta(t)$ and $\omega^2(t)$ to the leading order in $\alpha$
\be
\zeta(t)=\zeta_0(t) + \alpha \, \zeta_1(t) + {\cal O}(\alpha^2); \qquad \omega^2(t)= \omega_0^2(1-\frac{1}{2}\alpha \, t + { \cal O}(\alpha^2))
\ee
we can solve (\ref{Pinney}) perturbatively.
At the zeroth order, (\ref{Pinney}) leads to
\be
\DEE{\zeta}_0(t)+\omega_0^2 \, \zeta_0(t)=\omega_0^2 \, \zeta_0^{-3}(t)
\ee
which has the trivial solution $\zeta_0=1$, which is necessary for continuity at $t_0$. The first order equation is
\be
\DEE{\zeta}_1(t) + 4\omega^2_0 \, \zeta_1(t) - \frac{1}{2} \omega_0^2 \, t =0
\ee
with initial condition $\zeta_1(0)=0$, $\DE{\zeta}_1(0)=0$ and its solution is 
\be 
\zeta_1(t)=\frac{1}{8}\Big[ t- \frac{\sin(2\omega_0 t)}{2\omega_0}\Big]\; .
\ee
In the scaling limit we considered so far ($g_0\sim 1/L^2$) the function $\zeta(t)$, to the leading order in $\alpha$,  shows a cubic growth in time:  
\be
\zeta(t) \simeq 1+ \frac{\alpha}{12} \omega_0^2\,  t^3\; .
\ee
The quasi-adiabatic density profile (\ref{rhoEX}) can be characterized through the length scale
\be
\xi_{\text{qad}}(t) = \ell_0 (1+ \frac{\alpha}{12} \omega_0^2\,  t^3)= \ell_0 (1+ \frac{\alpha}{6} g_0\,  t^3)\; 
\ee
instead of the adiabatic length scale 
\be
\ell(t)= \frac{1}{(2g(t))^{1/4}}\simeq \ell_0(1+\frac{\alpha}{4} \, t)\; 
\ee 
identified in the previous section (see (\ref{rhoAD})). In the perturbative regime considered here, $g_0t^3\ll t$ and the expansion/contraction of the cloud is always slower than what would have been expected from a naive adiabatic guess.  
This feature is related to the freezing out of the dynamics close to the critical point with the consequent breakdown of the adiabatic behavior. 
In figure \ref{COMP} we show a comparison between the density profiles obtained from this Ermakov Lewis approach and the quasi-adiabatic approach developed in  the previous section. The comparison is made using $\varepsilon =\nu_g \, N \, \alpha \, t$, which follows from (\ref{epsilong}) and (\ref{g}), and we see an excellent agreement between both aproaches. 

\subsection{The special case of the decrease of the frequency as the inverse of time  }
A special case of interest is the situation where  the harmonic trap frequency decreases as the inverse of time
\be
\label{1t}
\omega(t)=\frac{\lambda}{t}
\ee
from an initial time $t_0=1$ set to one in the following such that the initial frequency is $\omega_0\equiv \omega(t_0)=\lambda$.
This time dependence is generated by the ramp 
\be
g(t)= \frac{1}{2}\left(\frac{\lambda}{t}\right)^2\; .
\ee 
In a recent work \cite{Shu} it has been shown that the release of a scale-invariant Fermi gas confined within a harmonic trap with this type of $1/t$ time dependence leads to the appearance of a discrete scaling symmetry in time. Such a discrete scale invariance is known to produce log-periodic modulations of the physical quantities \cite{logKarev}.  This behaviour has been observed and reported in \cite{Shu} where the size of the expanding Fermi gas grows through a sequence of plateaus which are distributed log-periodically. 
As the free Fermi gas is closely related to the Tonks Girardeau gas one expects that such a phenomenon also exist in that case \cite{Shu}. 
Indeed, here we prove that there is a regime where this log-periodic modulation of the expansion appears. 
Considering the Pinney equation with (\ref{1t})
\be
\frac{1}{\lambda^2} \DEE{\zeta}(t) + \frac{1}{t^2} \zeta(t) = \zeta^{-3}(t)
\ee
one can derive an equivalent Pinney equation with a time independent frequency. Indeed, with the substitution
\be
\zeta(t) = t^{1/2} r(\lambda \ln t)
\ee
we arrive at
\be
\label{pinney2}
r''(u) + (1-\frac{1}{4\lambda^2}) r(u)= r^{-3}(u)\; 
\ee
with boundary conditions
\be
\label{bc2}
r(0)=1\; , \quad r'(0)= -\frac{1}{2\lambda}\; .
\ee
Therefore, we have two distincts regimes, one with a high initial frequency $\omega_0=\lambda > 1/2$ for which 
\be
s^2 \equiv 1-\frac{1}{4\lambda^2} >0
\ee   
and an other one at low initial frequency, $\omega_0=\lambda < 1/2$, for which 
\be
-\kappa^2 \equiv 1-\frac{1}{4\lambda^2}  < 0 \; . 
\ee
\begin{figure}
\centering
\includegraphics[width=0.9\textwidth]{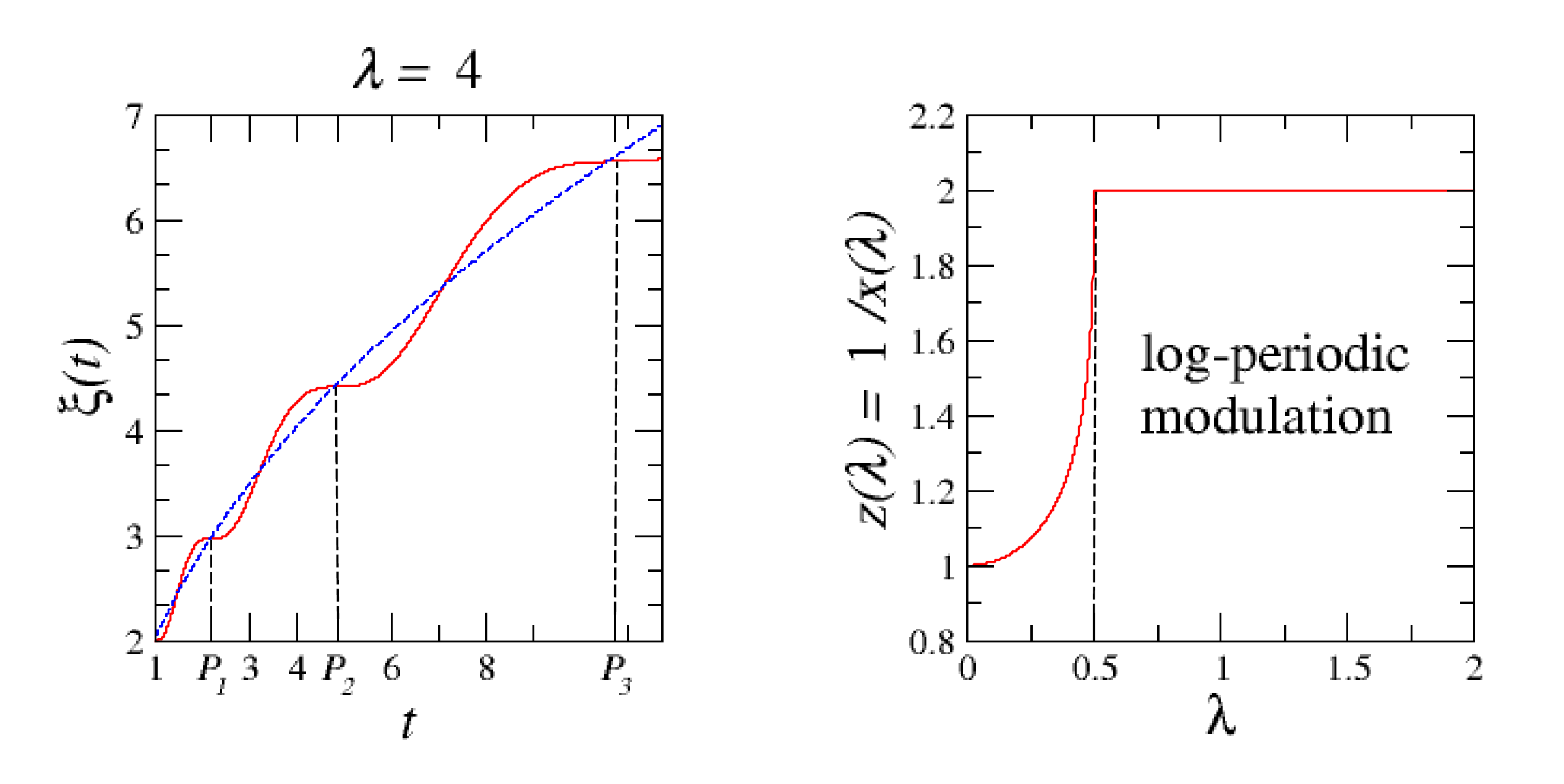}
\caption{\textsl{Left}. The solution of the Pinney equation for the protocol $\eqref{1t}$ at high initial frequency $\lambda>1/2$. It exhibits a square root dependence in time (blue dashed line) with a log-periodic modulation (red line). The solution exhibits a series of plateaus located at times $P_n=\exp(2n\pi/\sqrt{4\lambda^2-1})$. \textsl{Right.} The behavior of the dynamical exponent $z(\lambda)$ as function of the initial frequency: it varies continuosly from a ballistic value $z=1$ to a diffusive one $z=2$.}\label{efimov}
\end{figure} 
The solution of (\ref{pinney2}) with the boundary conditions (\ref{bc2}) is in the high initial frequency regime 
\be
\label{logperiodic}
r(u) = \frac{1}{s}\left[1-\sqrt{1-s^2} \sin \left( 2s u +\arcsin\sqrt{1-s^2}\right)\right]^{1/2}
\ee
and 
\be
\label{power}
r(u) = \frac{1}{\kappa}\left[-1+ \sqrt{1+\kappa^2} \cosh \left( 2\kappa u - {\rm arcosh}\, \sqrt{1+\kappa^2}\right)  \right]^{1/2}
\ee
in the low initial frequency one. The typical size of the bosonic cloud $2\ell_N(t)=2\sqrt{2N}\xi(t)$ is extracted from  the exact density profile (\ref{rhoEX}) and it is given through 
\be\label{xiEfi}
\xi(t)=\ell_0\, \zeta(t)= \sqrt{\frac{t}{\lambda}}\; r(\lambda \ln t)\; .
\ee
Clearly, we see in the high initial frequency case, that is for $\lambda >1/2$, a square root expansion with a typical log-periodic modulation. On the contrary, at lower initial frequencies, for $\lambda < 1/2$, the expansion of the cloud at long times is governed by a  pure power-law
\be
\xi(t) \sim t^{x(\lambda)}
\ee 
with a dynamical exponent $z(\lambda)=1/x(\lambda)$ that varies continuously with the initial frequency $\lambda$ and which is given through
\be\label{Xexp}
x(\lambda) = \frac{1}{2} + \kappa(\lambda) \lambda = \frac{1}{2}\left(1+ \sqrt{1-(2\lambda)^2}\right)\; .
\ee
Such a behaviour is reminiscent of marginal perturbations, such as the Hilhorst-van Leeuwen ones \cite{hvlKarev,TuPeIg,hvl}, affecting the equilibrium critical exponents at a second order phase transition continuously. 
Notice that the log-periodic modulation can be seen as the emergence of a complex  exponent $x(\lambda)$. Indeed, in the high frequency regime, the parameter $\kappa(\lambda>1/2) =i s(\lambda)$ and the power law behaviour 
$t^{1/2+ \kappa(\lambda) \lambda}= t^{1/2(1+i 2s(\lambda) \lambda)}$. Taking its real part gives the log-periodic modulation:
\be
\zeta(t) \sim \sqrt{\Re\{t^{1+2is(\lambda)\lambda}\} } \sim \sqrt{t \cos[2s(\lambda)\lambda\ln t]}\; .
\ee 
In figure \ref{efimov}, we report the typical length-scale $\eqref{xiEfi}$ for $\lambda>1/2$ and the behavior of the dynamical exponent $\eqref{Xexp}$ as a function of $\lambda$. 
\section{Summary and conclusions.}\label{CON}

We have investigated the unitary dynamics of a low-density one-dimensional gas of impenetrable bosons at zero temperature under a harmonic trap release.  In particular, we have studied  the off-equilibrium aspects emerging from a slow general time ramp of the trapping potential.  The presence of the trap leads to the emergence of a typical length scale $\ell(t)$ (see $\eqref{ell}$) which fully characterizes the time evolution of the cloud for an adiabatic process (see figure \ref{densAD}). 
We have computed the first off-equilibrium corrections arising away from adiabaticity using time-dependent perturbation theory. 
The departure from the equilibrium density profile  has been obtained and shown to exhibit a scaling form (\ref{Drho}) in the large particle number limit. We have also provided an exact solution using dynamical invariants in the case of a linear time ramp. In that case, we have identified the exact typical length scale $\xi(t)$ governing the process. This length scale is the product of the instantaneous initial typical length scale $\ell(t_0)$ with the solution $\zeta(t)$  of the Pinney non-linear differential equation $\eqref{Pinney}$. 
The connection with the first order perturbative result is made through a series expansion of the exact scale $\xi(t)$ in the limit of a small linear quench rate. Thanks to that the density profile, obtained from first order perturbation theory, matches perfectly the exact one.  
When the trap frequency decreases as the inverse of time, we have shown by an exact solution of the Pinney equation that the cloud expansion is modulated by a log-periodic function reminiscent of a discret scale invariance for high initial frequencies. If the initial frequency is low enough the expansion follows a power law with an exponent that continuously depends on the initial frequency. 
The proper identification of a typical length scale for these slow out-of-equilibrium processes may be useful in  experimental contexts
as for example in cold atoms setup. Indeed, the characteristic scales that we pointed out may give a theoretical hint on the fluctuations of the particle density due to smooth modification of the optical cavities.

\ack
We would like to thank J. Unterberger and J. Dubail for useful discussions. We  also gratefully acknowledge R. Qi for putting to our attention the nice work reported in \cite{Shu} who gave us the hint to consider the $1/t$ case treated in the last part of this work.

\section*{References}

\end{document}